# Automated Profile-Guided Replacement of Data Structures to Reduce Memory Allocation

Lukas Makor[a] , Sebastian Kloibhofer[a] , Peter Hofer[b] , David Leopoldseder[b] , and Hanspeter Mössenböck[a]

a Institute for System Software, Johannes Kepler University Linz, Austria
b Oracle Labs, Austria

**Abstract**    Data structures are a cornerstone of most modern programming languages. Whether they are provided via separate libraries, built into the language specification, or as part of the language's standard library—data structures such as lists, maps, sets, or arrays provide programmers with a large repertoire of tools to deal with data. Moreover, each kind of data structure typically comes with a variety of implementations that focus on scalability, memory efficiency, performance, thread-safety, or similar aspects.

Choosing the *right* data structure for a particular use case can be difficult or even impossible if the data structure is part of a framework over which the user has no control. It typically requires in-depth knowledge about the program and, in particular, about the usage of the data structure in question. However, it is usually not feasible for developers to obtain such information about programs in advance. Hence, it makes sense to look for a more automated way for optimizing data structures.

We present an approach to automatically replace data structures in Java applications. We use profiling to determine allocation-site-specific metrics about data structures and their usages, and then automatically replace their allocations with customized versions, focusing on memory efficiency. Our approach is integrated into GraalVM Native Image, an Ahead-of-Time compiler for Java applications.

By analyzing the generated data structure profiles, we show how standard benchmarks and microservice-based applications use data structures and demonstrate the impact of customized data structures on the memory usage of applications.

We conducted an evaluation on standard and microservice-based benchmarks, in which the memory usage was reduced by up to 13.85 % in benchmarks that make heavy use of data structures. While others are only slightly affected, we could still reduce the average memory usage by 1.63 % in standard benchmarks and by 2.94 % in microservice-based benchmarks.

We argue that our work demonstrates that choosing appropriate data structures can reduce the memory usage of applications. While we acknowledge that our approach does not provide benefits for all kinds of workloads, our work nevertheless shows how automated profiling and replacement can be used to optimize data structures in general. Hence, we argue that our work could pave the way for future optimizations of data structures.

## The Art, Science, and Engineering of Programming

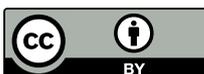

## 1  Introduction

Data structures and containers are ubiquitous in programs and programming languages [13]. Ranging from array abstractions and associative arrays to full-fledged collection and data structure frameworks, most programming languages come with a wide variety of container types to fulfill different purposes. Additionally, individual collection types may come with different implementations, tailored towards different criteria such as algorithmic complexity, memory efficiency, preserving insertion order, or sortability.

Numerous studies have shown the impact of using the right data structures, focusing on factors such as performance overhead and benefits [13, 31], memory utilization and bloat [10, 13, 19, 35], as well as energy efficiency [23, 34, 45]. The main cause of concern highlighted by these studies is frequently the right *selection* of data structure implementations [13, 19, 31, 35]. Various approaches tackle this challenge by using profiling information [5, 12, 33, 56, 70] or learning models [11, 15, 26, 31, 33] to automatically or semi-automatically identify inefficient usages and sometimes even correct them. However, many of these approaches either only give hints to the developer [56, 63], provide an abstraction of the data structures to enable on-demand swapping [15, 44, 70], or require dedicated test suites to enable the selection [5]. While such approaches can aid developers in their choice of implementations, they are often futile, when the affected data structures are not allocated within user code, but rather in frameworks or libraries on which developers have no control.

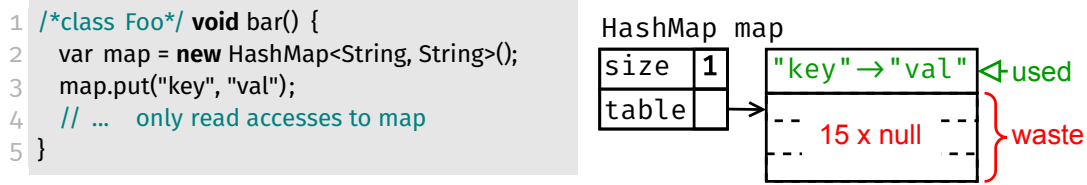

**(a)** HashMap allocation and element insertion.   **(b)** The memory layout of the HashMap.

■ **Figure 1**  Inefficiency of a java.util.HashMap when only containing a single entry.

The Java code in Figure 1a depicts the allocation of a java.util.HashMap within the method Foo.bar. Afterwards, a single key-value pair ("key", "val") is stored in the HashMap. This insertion triggers the allocation of the internal table storing the HashMap entries, which has a default initial size of 16 (in Java version 17). Hence, only a small part of the allocated table is actually used at this point, as illustrated in Figure 1b. In case no further elements are inserted into the HashMap, 15 of the 16 slots in the table would stay unused and waste memory. Therefore, using a custom initial size or even a HashMap implementation specialized for a single entry would be more memory efficient in this scenario. However, this requires manual developer intervention and knowledge about the number of items that will be stored in that HashMap. Whether developers can know in advance, for which purposes their data structures are used, is highly dependent on the program itself: If the map in this example is exposed to other parts of the program, it may be hard to identify *all* the different possible access locations to this map beforehand. Hence, manually performing such optimizations is typically not feasible for developers.

In this work, we demonstrate the automated replacement of data structures with more memory-efficient variants based on profiling information in GraalVM Native Image [64, 65], a system for *ahead-of-time (AOT)* compilation of Java applications. Apart from the execution of profiling runs (which are part of *profile-guided optimization (PGO)* [3, 8, 55, 64]—a feature of the underlying compiler infrastructure), our approach requires no user interaction and decides automatically whether to perform a replacement or not. We can replace data structures at individual allocation sites such that specific code locations that use inefficient data structures can be exploited and optimized, without affecting other uses of the same data structure type. The main goal of our data structure optimizations is to reduce the memory usage of data structures and thereby reduce the overall memory utilization of applications. Serverless computing—a well-suited area of application for GraalVM Native Image [65]—could therefore be a major beneficiary of this approach.

We showcase the viability of our approach by introducing several replacements for four data structures from the Java Collections Framework [40]: java.util.HashMap, java.util.LinkedHashMap, java.util.HashSet (internally backed by a HashMap), and java.-util.ArrayList. We chose these collection implementations, because studies have shown that they are among the most frequently used data structures in benchmarks and common Java programs [13, 35].

Overall, this work presents the following contributions:

1. An extension of GraalVM Native Image PGO for allocation-site-specific data structure profiling and for tracking of a variety of different data structure metrics.

2. A technique for automated replacement of data structure allocations during AOT compilation with more memory-efficient variants. Our replacements are not limited to user code but also affect third-party code such as frameworks or libraries.

3. An evaluation of the contained data structures and their properties in standard benchmarks and benchmarks based on microservice frameworks as well as an evaluation of the effects on memory allocation, performance, and binary size of our approach using this benchmark corpus.

In Section 2, we introduce the underlying AOT compiler infrastructure and discuss our integration into it. In Section 3, we outline our instrumentation and profiling approach and detail the metrics we collect for data structures. We describe our instrumentation framework and our replacement data structures in Section 4, where we also briefly discuss the composition of our benchmark corpus. A detailed explanation of our data structure replacement approach constitutes Section 5—we further describe techniques that we apply to reduce the memory usage of our replacement data structures. In Section 6, we discuss shortcomings of our approach. Section 7 contains a detailed evaluation of our approach and its effects on a variety of benchmarks. Finally, Section 8 describes related work.

## 2 Background

We integrated our approach into GraalVM Native Image [38]: We use its profile-guided optimization feature [3, 8, 55, 64] to first collect information about data structures in a profiling run on an instrumented executable and subsequently replace targets based on this profiling information during the compilation of the final executable. Apart from executing the profiling run, this process is fully automated and does not require human intervention.

## 2.1 GraalVM

GraalVM is a production-grade polyglot virtual machine written in Java [38, 68]. Besides supporting Java-bytecode-based languages such as Java, Kotlin, or Scala, it also allows the execution of a variety of other languages such as Python [37], Ruby [43], JavaScript [39], or C/C++ [50] via abstract syntax tree interpretation [21, 66, 68, 69]. GraalVM features a dynamic just-in-time (JIT) compiler that uses run-time profiling information for optimization [16, 29, 59]. Internally, the GraalVM compiler uses a graph-based sea-of-nodes IR (Graal IR) [16, 17] and applies a variety of optimizations on a method's Graal IR, before emitting machine code. The approach presented in this work is based on GraalVM 23.0 with Java version 17.

## 2.2 GraalVM Native Image

Our approach is based on GraalVM Native Image [64, 65], a component within GraalVM that enables *ahead-of-time* (AOT) compilation of applications. By using a *points-to* analysis [24, 53, 58] to identify reachable types and methods, GraalVM Native Image compiles programs under a closed-world assumption to a native binary (the so-called *image*). This binary subsequently no longer requires a JVM for execution. The *SubstrateVM* virtual machine implementation is built into the binary and provides services such as garbage collection and thread management [42].

A major goal of GraalVM Native Image is to significantly reduce the startup overhead of typical JVMs, caused by the initial warm-up of the virtual machine and the program itself (which initially runs in the interpreter), and to lower the memory footprint of applications [65]. Support for certain Java features such as run-time reflection or dynamic class loading is limited and has to be configured by the developer [65].

**Profile-Guided Optimization**   GraalVM Native Image relies on the GraalVM compiler to optimize methods after points-to analysis, but unlike with JIT compilation, no profiling information about the program is readily available to guide optimization decisions [65]. Instead, Native Image supports profile-guided optimization [3, 8, 46, 55, 64] in two steps. PGO first produces an *instrumented* image of the target program which is subsequently executed. During this run, the instrumented image collects profiling information on types, method calls, conditionals, etc. and writes this information to a file as the program terminates. A subsequent build of the image then uses the gathered profiling information to perform more advanced optimizations, comparable to those applied by a JIT compiler.

The primary goal of Native Image's PGO is to enable more sophisticated optimizations in an AOT compiler by using information about the run-time behavior of the corresponding application, akin to a JIT compiler. Just as a JIT compiler collects profiling information about the program *being executed*, an AOT compiler collects profiling information about the program *to be executed*. In the optimium case, the profiling exactly matches the real program input so that the compiler can "overfit" the optimizations to the target program. This is similar to how profiling is used in a JIT compiler. However, while a JIT compiler always profiles and optimizes the current program execution, Native Image collects PGO profiles for a certain workload ahead of compilation time and can thus optimize programs that reflect this workload. For different workloads a new PGO profile has to be collected. This approach is also used in related work on PGO [9, 20, 28, 32, 60, 67, 73]. It must not be confused with machine-learning-based profiling and optimizations, which typically uses a far larger and more varied set of inputs to perform optimizations on a wider range of applications, achieved via sophisticated machine learning techniques [52]. Their goal is training a model that can reliably predict optimization potential in new programs based on a large set of training inputs, whereas with PGO, we want to optimize the program towards a given workload. Creating a generalized version of the program that performs fast on many different inputs is an explicit non-goal of this kind of PGO. Our evaluation is therefore conducted similarly. We use the same workload for profiling and for measurements, although the size of the workload is smaller for profiling than for the measurements.

Typical downsides of PGO are the comparatively complicated setup (two different compilations) and the relatively high overhead during the instrumented execution. However, the latter is performed offline, hence it does not contribute to the execution time in production but rather slows down the build process.

For our approach, we first profile the use of certain target data structures and then use PGO to optimize and replace them in a second build step. For the remainder of this work, we will refer to the image that is generated to collect profiling information as the *instrumented image* and to the image that was created *using* the gathered profiling information as the *optimized image*.

## 3 Allocation-Site Profiling

Native Image's PGO instrumentation introduces atomic counters into the compiled image that accumulate metrics in a pre-allocated off-heap memory region. Our goal is to replace data structure allocations at individual allocation sites. At each allocation site in the code, we want to pinpoint how the data structure objects allocated there behaved at run time, what operations where performed on them, and track other properties such as the contained elements or the data structure growth. Hence, we want to extend this instrumentation to generate allocation-site-specific profiles and track custom, data-structure-specific metrics.

### 3.1 Profiling Metrics

Currently, we instrument HashMap and HashMap (subtype of HashMap), HashSet, and ArrayList instances. Table 1 summarizes the most important metrics that we track for each data structure:

■ **Table 1** Overview of all data structure metrics we profile.

| Metric | Type[a] | Description |
|--------|---------|-------------|
| allocations | *MSL* | The number of allocations at a specific allocation site. |
| size | *MSL* | The maximum size any instance allocated at this site reaches. |
| size class ⟨**x**⟩ | *MSL* | A size class counts the number of data structures allocated at a particular allocation size, with their maximum element count falling within the corresponding size class boundaries. Currently, we track size classes with bounds (⟨**x**⟩) 0, 1, 2, 8, 16, 64, 256, 1024, 65536, and "inf" (exceeding all other size classes), where the bound defines the (inclusive) size limit of the size class. For example, a data structure with a maximum size of 17 counts towards *size class 64*. |
| gets | *M* | The number of read accesses (`Map.get` calls) on Maps allocated at this site. |
| inserts | *M* | The number of (successful) element inserts on data structures allocated at this site. |
| element type | *L* | Encoding of the element types of ArrayLists allocated at this site. We distinguish between 8 Java primitive types (`byte`, `short`, `int`, `long`, `float`, `double`, `char`, `boolean`) and objects. |
| entry access | *M* | The number of times a map entry is exposed via `Iterator.next` calls on a map's entry set. |

[a] M ... (Linked)HashMap, S ... HashSet, L ... ArrayList

**Size and Size Classes**     As we are interested in the memory utilization of data structures, we track the maximum size of data structures allocated at a specific allocation site. We further track the maximum size class that individual data structures reach. Therefore, after each operation affecting the size of a data structure, we check whether the current data structure size exceeds the boundary of the maximum size class for its allocation site and update the size class accordingly. We use power-of-two size classes since the size of the internal table of a HashMap (and therefore a HashSet) is also doubled upon resizing. The size class selection is based on thresholds that cover both a variety of smaller data structure sizes as well as larger ones that contribute more to overall memory usage. While the maximum size is useful to get an overall estimate of how large data structures allocated at a specific allocation site may get, the size classes enable a more nuanced assessment: If at a certain allocation site, most data structures stay empty, while one instance grows to hundreds of elements, the maximum size would only tell as the latter fact, while disregarding the many empty data structures. When using size classes, we get exact counts for each size class that tell us, how many instances at this specific allocation site reached a certain maximum size. This is important for our replacement heuristics described later in Section 4.

**HashMap Entry Accesses**     HashMaps store their data in so-called `Entry` objects that contain the key, value, and the key's hash code. These entry objects further act as list nodes in case of collisions. While these objects are data structure internals, they actually can be exposed by iterating over a map's *entry set*, i.e., a view of the set of Entry nodes within a HashMap defined by the `java.util.Map` interface.

One alternative to such an "entry-based" implementation for a HashMap would be some variant of an ArrayMap [1], where the data is directly stored—as the name suggests—in internal arrays instead. This has the benefit of requiring no "intermediary objects" such as HashMap's *entries*, hence ArrayMaps are typically more memory-efficient, albeit with a slight drop in performance [1]. In compliance with the java.util.Map interface, however, even such an implementation would have to provide some means to get access to an entry set. Hence, when using the entry set of the map, Entry objects need to be created nonetheless. This problem is also outlined by other popular Java map implementations [2, 18, 25, 30]. Hence, we consider this metric vital for memory considerations. We will come back to this point when introducing our replacement data structures and their corresponding heuristics in Section 4.

**HashMap Inserts**   While the size metrics let us estimate the overall memory usage of a data structure, they are insufficient to assess its run-time behavior. Due to the aforementioned internal *entry* objects that HashMap uses for data storage, inserting new elements always causes additional allocations. Because HashMaps may grow and shrink over time (thus affecting the size metrics), tracking the total number of inserts allows us to estimate how many entry objects are in fact created by inserting elements into HashMaps allocated at a specific site.

**ArrayList Element Types**   Finally, when storing primitive values, Java's ArrayList uses *boxing*, i.e., wrapping the primitive value into an object, thus using more memory than necessary to store the actual data. Hence, we track the types of elements added into ArrayLists per allocation site to specifically detect instances in which only primitive values of a single type are stored.

### 3.2 Instrumentation via Allocation-Site-Specific Counters

As it is a non-trivial problem to identify the allocation site of a specific object at run time (e.g., the object may have been created within another method or the reference may *alias* with other references), we have to track an allocation-site-specific offset into the profiling region within the actual object. In the profiling run, we therefore add an additional field, used to store that offset, to each profiled object type. We calculate this offset at compile time after method inlining. Then, most constructors are embedded into their corresponding callers. The offset is then based on the position of the constructor call within the caller and is thus unique for this particular compilation. If the same constructor is inlined into another method, a different offset is calculated, thus resulting in independent profiles. We consider the inlined location as the actual allocation site of the object for the rest of the paper. At each instrumented method within a data structure, we then accumulate counters using this offset. Hence, all objects originating from a specific allocation site access and modify the same counters.

The constructor call in Figure 2a is automatically altered to also generate this allocation site's unique profiling offset and store it in the injected field $profilingOffset of the allocated HashMap. Figure 2b depicts the substitution and instrumentation of the HashMap.afterNodeInsertion method (an empty method in HashMap that is only overwritten in subclass LinkedHashMap). In Line 4, we increment the insertion count,

```
1  /*class Foo*/ void bar() {
2    // allocation site has offset <o>
3    var map = new HashMap<...>();
4    // automatically store offset
5    map.$profilingOffset = <o>;
6    // ...
7  }
```

```
1  void afterNodeInsertion(boolean evict) {
2    var curSize = this.size;
3    var prevSize = curSize - 1;
4    _count(INSERT, $profilingOffset);
5    _max(SIZE, $profilingOffset);
6    _sizeClass($profilingOffset, prevSize, curSize);
7  }
```

**(a)** Instrumentation of the HashMap allocation site.

**(b)** Profiling of the insertion count, max size, and size class via HashMap.afterNodeInsertion.

■ **Figure 2** Instrumentation of java.util.HashMap: Injecting and assigning an allocation-site-specific offset and tracking metrics upon interaction.

while Line 5 tracks the current maximum size of all HashMaps allocated at the same site. As the insertion of a new element may affect the *size class* the map is currently associated with, we also have to check for that (Line 6).

### 3.3 Profiling Format

During the profiling run, Native Image PGO keeps the profiling information in memory and only serializes the collected information into a JSON file upon program shutdown.

The format of such a profile file is shown in Listing 1: Line 1 denotes the allocation site (i.e. the method and bytecode index—typically encoded for compression) within the introductory example method from Figure 1. The array in Line 2 contains the profiling data. The record in Line 3 shows the profile if some calls to Foo.bar are inlined into a method Foo.baz. This nested context later allows us to separately replace either allocation with distinct alternative implementations (e.g., a map specialized for size 1—cf. Section 4). This file is subsequently loaded when building the optimized image. With the ctx, the profiling information can be re-associated with the corresponding allocation site to then perform optimizations, which in our case means replacing the data structure with a better version if possible.

■ **Listing 1** Format of profiling entries.

```
1  {   "ctx": "Foo.bar(): 4",
2      "records": [ 70 /*allocations*/, 1 /*max size*/, 76 /*get calls */, ... ]
3  }, { "ctx": "Foo.bar(): 10 > Foo.baz(): 4",
4      "records": [ 14 /*allocations*/, 10 /*max size*/, ... ]
5  }
```

## 4 Data Structure Analysis and Optimization

Based on the instrumentation process summarized in Section 3.2, we conducted an analysis of the size classes in which data structures occur. This analysis was performed on a benchmark corpus consisting of both standard benchmark suites (*dacapo 9.12-MR1-bach* [6], *scala-dacapo 0.1.0* [54], *renaissance 0.9.0* [47]) and benchmark suites used for Native Image evaluation, which are based on the popular web and microservice frameworks *Spring* [62], *Quarkus* [48], and *Micronaut* [51]: Each microservice benchmark suite contains a so-called *helloworld* benchmark that is based on the corresponding framework's laucher or "getting started" guide, i.e. the basic use cases of each framework. Furthermore, each suite contains one additional more complex benchmark, parameterized with different workload sizes (*tiny*, *small*, *medium*, etc.). The *Spring* suite contains an adaptation of Spring Boot's PetClinic Sample Application [61]. *Quarkus* further contains a microservice benchmark based on Quarkus' Apache Tika Quickstart application [49]. *Micronaut*'s second benchmark is a web shopping implementation called *ShopCart*.

The results of our size class profiling for these benchmark suites are presented in Figure 3. Each bar represents the number of objects of the respective type that reached the corresponding size class. First, it shows that while all data structures are frequent in most benchmarks, scala-dacapo generally shows the fewest occurrences, particularly for HashMaps and HashSets. It further shows that large HashMaps and HashSets only appear in the dacapo and renaissance benchmark suites. Conversely, almost all benchmark suites show a prevalence of small or even empty data structures—a fact that we exploit for our data structure replacements.

The next section details our specialized data structure variants based on this analysis that we insert upon replacement.

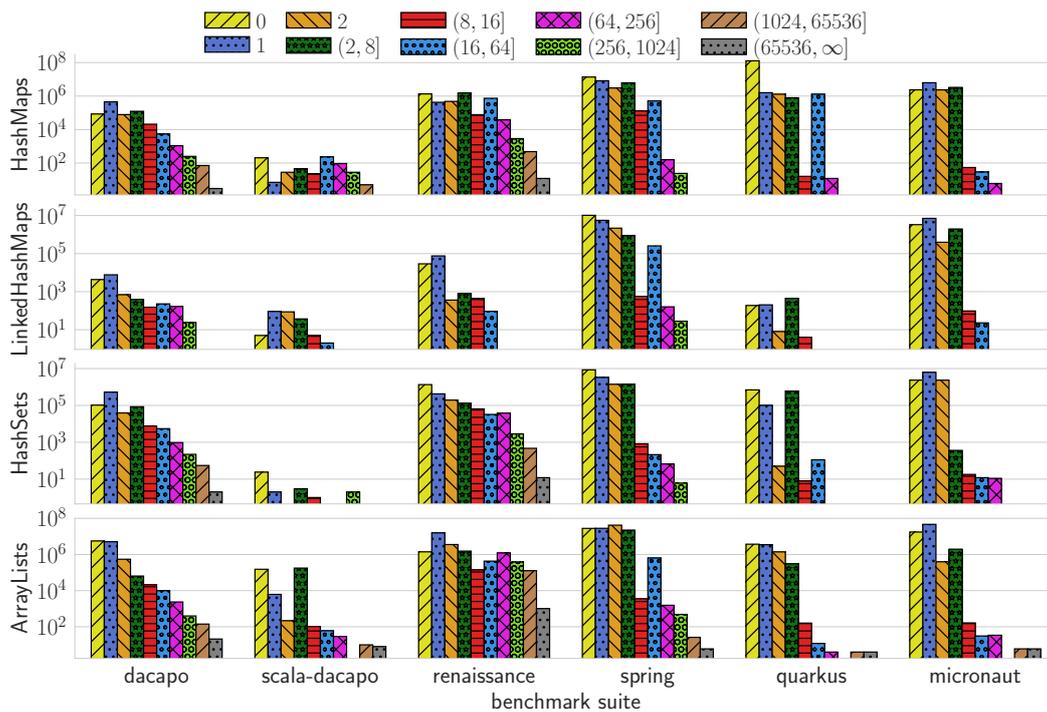

■ **Figure 3** The largest reached size classes of data structures allocated in common benchmark suites (logarithmic scale).

**Optimized Data Structures**   As our profiling is performed per allocation site, we want to replace the original data structure allocations with allocations of an optimized data structure implementation, when the allocation site exhibits homogeneous and optimizable characteristics. Our replacement types are *subclasses* of the corresponding original type so they are compliant with the original API.

**Fixed-Size Specializations**   Our analysis showed that many allocated data structure instances only reach maximum sizes of 0, 1 or 2 elements. This insight is backed by similar observations made by Chis et al. [10]. By default, data structures such as ArrayLists and HashMaps internally store their data using arrays. However, when the maximum size is known and small enough (specifically in the cases of 0, 1 and 2 elements), storing the elements directly in fields instead of allocating an array is more efficient. Therefore, we implemented versions of each data structure specialized for sizes 0, 1 and 2. We use the corresponding *size class* information (size class 0, 1 or 2) to decide upon replacing an allocation with such a specialized version. This decision is based on a threshold of 95 %, i.e., at least 95 % of the data structures allocated at that allocation site must fall into the specific or a smaller size class. The reason for this threshold is that in case the profiled size is incorrect the specialized versions incur a small overhead in performance and memory. Therefore, we want to avoid those cases without limiting the optimization potential too much.

As our HashMap implementations do not store `Entry` objects, but need to support the full HashMap API, we utilize another constraint when replacing HashMap allocations. To limit the number of `Entry` allocations that need to be performed by our memory-efficient implementations to a reasonable number, we only perform the replacement when the number of entry accesses is less than 5 % of the number of inserts at this allocation site. For the fixed-size implementations, we allow replacement as long as the number of entry accesses is less than 3, as we still achieve memory improvements up to that threshold. The reason for these constraints is that our HashMap implementations save memory when inserting elements, but allocate additional memory on each entry access. Hence, we ensure that allocation sites of HashMaps that exhibit frequent accesses to entry objects are not replaced.

As profiling data is not guaranteed to reflect the actual program behavior (e.g., in production, the allocated data structures may become larger or behave differently), we also have to provide a *fallback* in case the profile does not match. Therefore, we add a field for the fallback data structure to each of our fixed-size replacement types. In case the requirements of a replacement type do not hold anymore, an instance of the original type (that was replaced with the replacement type) is allocated and all existing entries are moved to the fallback data structure, which also is utilized for all further operations.

Listing 2 shows parts of the implementation of `SingletonHashMap`, a HashMap replacement that is efficient if the HashMap contains only a single key-value pair. The map has 4 fields that store the single key and value pair, the state of the map (due to HashMap allowing null keys, we cannot distinguish between an empty map and a single null entry based on the other fields), and the fallback data structure. As shown in `SingletonHashMap.put`, based on the current state of the map, it either traverses a fast path where the individual key-value pair is set/modified or initializes the fallback data structure (Line 14) if more than one key-value pair is inserted. Later invocations of the method subsequently directly use the fallback (Line 15). `SingletonHashMap.initFallback` may also be invoked by other methods that modify the map and may have to fall back to the original implementation. Hence, if the replacement based on the profiling turns out to be ill-advised, two kinds of costs occur: First and immediately, the fallback data structure has to be initialized and existing elements have to be added. Subsequent accesses then use the fallback data structure, hence, one additional method call per data structure access has to be performed. The other fixed-size specializations (also for other data structures) are implemented similarly.

■ **Listing 2**  Excerpt of the implementation of SingletonHashMap and its use of the fallback mechanism when inserting new entries.

```
1  class SingletonHashMap<K, V> extends HashMap<K, V> {
2      K cachedKey = null;          // the single cached key (if any)
3      V cachedValue = null;        // the single cached value (if any)
4      byte state = 0;              // empty (0), singleton (1), fallback (2)
5      HashMap<K, V> fallback = null; // the fallback map
6      V put(K key, V value) {
7          if (state == 0) {  // fast path: insert into empty map
8              cachedKey = key; cachedValue = value; state = 1;
9              return null;
10         } else if (state == 1) {
11             if (Objects.equals(cachedKey, key)) { // fast path: overwrite existing key
12                 V temp = cachedValue; cachedValue = value;
13                 return temp;
14             } initFallback(); // slow path: allocate fallback map + add any cached elements
15         } return fallback.put(key, value); // slow path: map is in fallback mode
16     }
17     void initFallback() {
18         fallback = new HashMap<>();
19         if (state == 1) { // the map already contained a cached element
20             fallback.put(cachedKey, cachedValue);
21             cachedKey = cachedValue = null;
22         } state = 2; // signals use of the fallback data structure
23 }}
```

**General Memory-Efficient Implementations**  In order to reduce the memory usage of HashMaps that are relatively large, but do not exhibit excessive get accesses, we utilize a map that, in contrast to HashMap, does not allocate intermediary *entry* objects, but rather uses a single array to store keys and values alternatingly. This implementation is based on the EconomicMap [36] used in GraalVM. In addition to the constraint regarding entry accesses that was explained in the previous section, we also require that 90 % of the HashMaps allocated at an allocation site need to have a max size of at least 5. Replacing smaller maps would not yield a significant improvement in terms of memory usage. Similarly, EconomicMap is optimized for maps with less than 256 elements, as larger maps require a larger hash size and impact both performance and memory usage. Therefore, only allocation sites with a max size below this threshold are considered for replacement with this particular map implementation. Furthermore, the average number of *gets* per allocated HashMap needs to be lower than the maximum size of the HashMaps allocated at that allocation site. This constraint was added to limit the performance impact in proportion to the saved memory. We do not alter the constructor arguments of the replaced EconomicMaps, so any provided value for the initial size stays the same.

The implementation of HashSet is inefficient as it uses a HashMap internally, where the *key* of each key-value pair is the actual HashSet data and the *value* is a singleton dummy object. Therefore, we utilize a more memory-efficient HashSet implementation based on open addressing and linear probing, which uses an object array to store the added objects. Consequently, this implementation does not require auxiliary objects for the data storage and is, therefore, more memory-efficient than HashSet. Hence, we replace all HashSet allocations that are not already replaced with one of the fixed-size replacement types with this implementation.

As we also track the types of the elements inserted into an ArrayList, we created ArrayList implementations specialized to a certain primitive type, i.e., the version specialized to int utilizes a primitive int array to store its values. Hence, when we detect that all elements inserted into an ArrayList are of the same primitive type, we utilize the implementation specialized for that type instead. Especially for smaller primitive types, e.g., boolean or byte, the per-element memory savings are significant. However, just like the fixed-size replacement types, these primitive type replacements are applied based on profiling data and cannot handle arbitrary operations. Hence, we support the fallback to a normal ArrayList if the element type restriction is violated at run time. Note that all of the thresholds mentioned above are configurable via command line options. The defaults are based on our observations during evaluation.

## 5 Profile-Based Data Structure Replacement

Section 3 introduced our instrumentation process and the different metrics we collect per allocation site. We devised an approach that automatically replaces certain allocations based on the gathered profiling information using the specialized data structures presented in Section 4. This process is integrated into GraalVM Native Image's PGO infrastructure (cf. Section 2.2). It first creates an instrumented binary from a program and subsequently generates an optimized version based on profiling information gathered before. In this section, we detail the replacement step that we integrated into the generation of the optimized image.

### 5.1 Allocation Site Replacement

As mentioned before, GraalVM Native Image features a powerful points-to analysis that incrementally detects reachable types and methods. In the process, the initial Graal IR of methods is created. We want to perform replacement *before* a method is analyzed, such that the analysis directly processes our replacement type instead of the original type. Hence, we implemented our replacements within a new compiler phase for GraalVM Native Image that is injected early on in the compilation pipeline. Similar to the instrumentation process, this phase iterates over the Graal IR and detects allocations of target data structures. We want to replace those allocations, where the profiling information hints at possible improvements, hence we only consider allocations for which profiling information is available and where the heuristics suggest a better alternative (cf. Section 4).

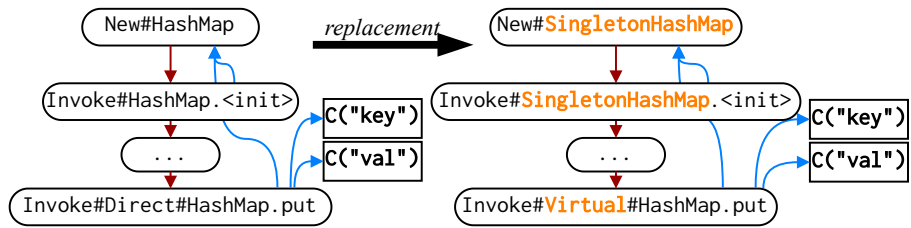

■ **Figure 4** Replacement of a HashMap allocation at compile time in Graal IR.

The left-hand side of Figure 4 depicts a simplified version of the Graal IR originating from the code of Figure 1a: At the early stages of the compilation, before inlining occurs and the IR nodes are transformed into more low-level operations, initializing an object consists of two parts, the allocation and the constructor call. The red arrows show the control flow—the allocation (New#Hashmap), the constructor invocation (Invoke#HashMap.<init>), and the subsequent put call (Invoke#HashMap.put)—, while blue arrows denote data flow dependencies. To safely replace the creation of a data structure with a custom type, we have to adapt these nodes. For the allocation, we simply replace the node with a copy that allocates the replacement type. Our replacement types are designed to have the same constructors as the original types. Hence, the corresponding constructor invocation node is replaced with its counterpart in the replacement type, followed by rewiring any arguments to the new invocation node. The put call in the original IR was interpreted as a so-called "direct" invoke, i.e., a non-polymorphic method call where the call target is known at compile time (the put implementation in java.util.HashMap) such that no dynamic dispatch is required. The replacement of the allocation complicates this process, however, as we would want to call the put method of our SingletonHashMap instead. Therefore, we also adapt all attached method calls and turn them into polymorphic ("virtual") calls. After this phase, the points-to analysis ensures that the new type information is propagated to other usages. While this transformation into virtual calls can impact performance either directly or by preventing other optimizations such as inlining, the evaluation of the performance in Appendix E overall shows only a minor dent in performance (0.60 % on average) with our approach. We therefore conclude that the impact of this process is negligible. This process allows us to target and replace individual allocation sites without suffering from performance penalties due to further indirections and method calls that wrappers or further abstractions would introduce.

### 5.2 Footprint Reduction for Replacements

Shacham et al. [56] discussed the costs and benefits of different replacement approaches. Our choice of using subclassing for our replacement data structures has the prime benefit of requiring as few modifications to the code as possible, while also simplifying the preservation of the public API of the affected data structures.

However, the usual drawback of this approach is that by inheriting from the original data structure type, each replacement type also inherits all the *fields* of the super type (unused in the replacement type). In the case of HashMap, for example, this concerns the original field for the entry array (albeit null and unused in our replacement data structures) as well as auxiliary fields used for caching the different views of the map (the values, the entry set, the key set). Specifically for the size-0, size-1 specializations (cf. Section 4), this would mostly nullify all the memory savings achieved by the specific implementation. However, as Native Image has to calculate and encode the object layout for each type in advance, we detect fixed-size replacement types during this process and simply *delete* the inherited fields from those objects. This has the interesting result that objects of the replacement type are subsequently *smaller* than objects of the parent class.

Note that this process does not impact compatibility of our approach: The targeted types only contain (package-)private and protected fields and our specialized implementations also overwrite (or replace with compiler support) all methods of the original classes that use those deleted fields. Hence, our implementations contain no usages of those fields; they only access their own declared fields.

## 6  Limitations

In this section, we discuss certain issues of our approach, which affect the program semantics after replacement as well as limitations of the *chosen* replacement technique.

**Semantics-Preserving Replacement**   Since unobtrusiveness is pivotal for our approach, we strive for preserving program semantics. While we fully support all public methods of data structures, there are some features that are not yet supported by our approach:

In Java, each object provides the `getClass` method that enables access to the object's type instance (a `java.lang.Class` object). In our current approach, invoking `getClass` on a replaced data structure actually yields the `Class` of the replacement type. The `getClass` method cannot be overwritten in Java. Hence, this is not something we can address simply by subclassing. We also do not want to modify `Class` objects, as they are used for type checking, contain a type's *virtual method table (vtable)*, and may also receive special handling by the compiler or the runtime itself.

The Class objects are also the cornerstone of Java's run-time reflection system that enables fine-grained class introspection. Java reflection is also a challenge for GraalVM Native Image, as method calls and field accesses performed via reflection often cannot be resolved during build time. Hence, the usage of reflection in Native Image typically has to be specifically marked via configuration files [65]. While this may seem like a major limitation, framework vendors have adapted to this by relying increasingly on code generation and build-time introspection instead of run-time reflection [27, 57]. The same limitation applies to our approach: As reflection enables direct access to the underlying Class object—and thus to all its fields and methods—it allows the user to access fields and methods added by the replacement type or to access fields of the original type that are not used in the replacement or even deleted (cf. Section 5.2).

One idea to tackle both, the problem regarding the Class object and reflection, is to disable replacement if we detect `getClass` calls or reflection usage on target data structures at build time. Despite these concessions, in our evaluation, we only encountered one single case in `java.util.stream.Collectors.toUnmodifiableList`, where the Class mismatch resulted in a run-time exception. Hence, we excluded that allocation site from replacement manually.

Various researchers have pointed out solutions to preserving the program semantics, such as fully replacing the corresponding Java Collection type to then wrap and—on-demand—change the underlying implementation [44, 56, 70], or introducing an extended static analysis or reasoning step to *only* perform replacement if compliance is assured [14, 63]. Naturally, our approach being integrated into an ahead-of-time compiler infrastructure brings both benefits and challenges in that regard: Native Image works under a closed-world assumption. Hence, whole program analysis is certainly *more* feasible than when having to deal with JIT-compiled Java code and dynamic class loading. However, the AOT compilation aspect also requires all replacements and optimizations to be done up front, using the information that is available. Hence, unlike in JIT compilers, there is no opportunity to later correct false assumptions—in our case revert replacements. As explained in Section 4, we use fallback data structures in such cases.

There is one other case where our replacements could violate program semantics: Our primitive ArrayLists store the values in primitive arrays. Hence, the values need to be unboxed when stored and boxed when returned. Therefore, the reference inserted is not guaranteed to be the same as the reference that is returned (even though the primitive value is the same). However, relying on reference equality of boxed values is generally discouraged [41]. So, we argue that this is a reasonable limitation.

**Context-Sensitive Profiling**    Inherently, Native Image's PGO works at a method level and generally does not consider context for method calls. As described in Section 3, we rely on inlining to get allocation-site-specific profiles. Nevertheless, even after inlining, we lack deeper allocation site call stacks. This lack of a deeper call stack could lead to some missed opportunities, e.g., a (not inlined) factory method could be used multiple times with vastly different usage patterns, thus resulting in a single profile, encoding all those behaviors. However, the results show that even limited contextual information leads to meaningful profiles and subsequently effective replacements.

## 7 Evaluation

We evaluated our approach on the benchmark corpus presented in Section 4, which consists of both standard benchmarks and custom microservice benchmarks. We executed the standard benchmarks with each benchmark's default input configuration and the microservice benchmarks with up to 5 different workload sizes. To obtain profiling data, we first execute each benchmark with a slightly reduced workload and then do the actual measurement run (cf. Section 2.2).

**Motivation of the Evaluation Metric**    Since our work focuses on reducing the allocated memory in applications, we measure the *total allocated bytes* per benchmark by aggregating the newly allocated objects and their sizes in each GC run. We acknowledge that *resident set size (RSS)* is another metric commonly used to measure application memory usage. However, we found that RSS is highly dependent on the used garbage collector and its policies for collecting and releasing memory. In our measurements, this led to RSS fluctuations of up to 50 % in many benchmarks. Thus, we argue that these fluctuations make it hard to estimate the impact of our optimizations on RSS. Therefore, we excluded the RSS results from the evaluation.

However, the issue still remains that the total allocated memory does not give us information about how long the replaced data structures live and, therefore, whether they have a lasting impact on the application's memory usage. Therefore, we further conducted a lifespan analysis of replaced data structures: There, we track the number of and memory occupied by replaced data structures that the garbage collector promotes to the *old generation*. The old generation contains long-lived objects that survive multiple garbage collections. Hence, this evaluation tells us whether our replacements impact potentially significant objects that have a prolonged impact on the currently allocated memory due to them residing longer in the heap. While we present the detailed breakdown per benchmark in Appendix C, the evaluation showed that for standard benchmarks out of all bytes of objects that are promoted to the old generation only 3.6 % are contributed by data structures. For microservice benchmarks the ratio is slightly higher with 4.7 %. We further investigated, how many of the data structure instances within the old generation could be replaced: In standard benchmarks, the ratio of replaced data structure instances in the old generation is 50.7 %, while in microservice benchmarks it is 59.4 %. Therefore, we conclude that despite the small impact of data structures in general, our approach can target a large portion of those that remain longer in memory.

**Benchmark Methodology**   We compare the execution of the benchmarks using the standard version of GraalVM Native Image with their execution using our modified version of Native Image, where data structures are automatically replaced. In both executions, each benchmark is run once before the actual execution to capture its profile for PGO (and our replacements). A replacement in this context means the replacement of all allocations at a specific allocation site. Hence, one replacement can lead to multiple allocations of the replaced data structure. In Appendix A, we analyze the replacements by summarizing the replaced allocation sites and the data structures allocated there for each benchmark. All experiments were executed on an Intel I7-4790K @ 4.4 GHz with 20 G main memory. Hyper-threading, frequency scaling, and network access were disabled. We used Java version 17 for the evaluation.

### 7.1 Data Structures in Benchmarks

First, we analyzed how our targeted data structures, i.e., (Linked)HashMaps, HashSets, and ArrayLists, contribute to the overall memory allocation in the benchmark corpus. For this calculation, we consider all "static" costs of the data structures, i.e., the data structure object and auxiliary objects (entries, table). Temporarily created objects such as a HashMap's *entry set* (see Section 3.1) that are allocated at certain accesses are not considered. Also, the actually stored data is not considered, as our optimizations typically do not affect the elements themselves but rather the way in which they are stored within the data structure. They are, however, included in the total allocated bytes as part of the evaluation in Section 7.2.

Figure 5 shows how our target data structures contribute to the total allocated bytes. It shows that those data structures play only a minor role in terms of memory allocation in most of our benchmarks (particularly scala-dacapo benchmarks show low percentages). However, there are some benchmarks where those data structures contribute significantly to the allocated memory (e.g., *dacapo:pmd*, *renaissance:fj-kmeans* and *(par-)mnemonics*, some microservice-based benchmarks). These are the benchmarks that represent the targets of our optimizations.

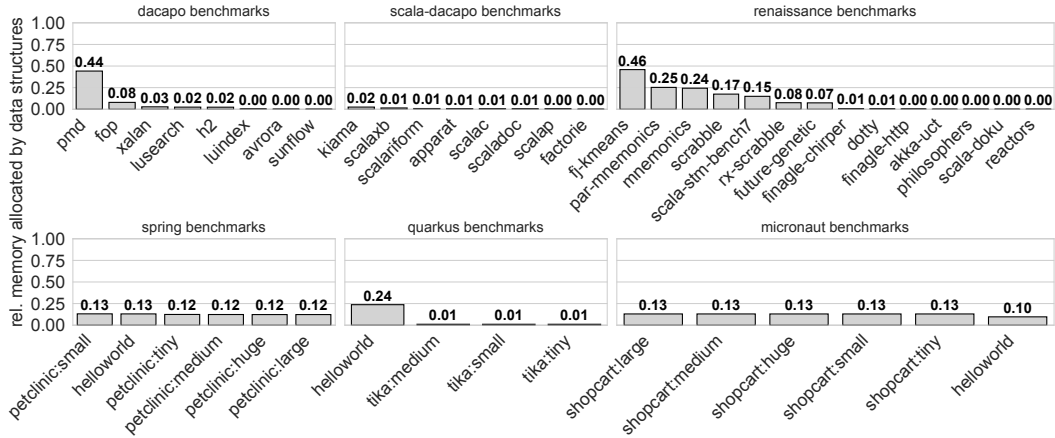

■ **Figure 5**  Relative contribution of our targeted data structures to the total allocated bytes in our benchmark set (ordered by contribution).

### 7.2 Impact on Memory Allocation

Figure 6 presents the results obtained with standard benchmarks (dacapo, scala-dacapo, renaissance). The chart shows the total allocated bytes per benchmark. With our replacements, we were able to lower the allocated bytes by 1.63 % on average (geometric mean). For dacapo, this means a reduction of 2.42 %, for scala-dacapo of 0.94 %, and for renaissance of 1.58 %. The small effect on scala-dacapo is to be expected, as the benchmarks of this suite have only few usages of our target data structures (cf. Section 4 and Section 7.1). Benchmarks such as *dacapo:pmd*, *renaissance:scala-stm-bench7*, and *mnemonics* benefit the most with reductions of 13.85 %, 10.32 %, and 3.28 %, respectively. In Appendix B, we show a detailed overview of the corresponding reductions in allocated bytes per data structure. The most prominent results are summarized in the following paragraphs.

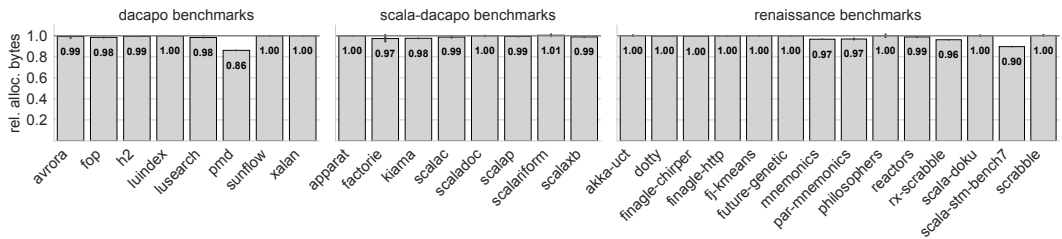

■ **Figure 6**  Evaluation of our approach on standard benchmarks, showing the ratio of the total allocated bytes within each benchmark (*lower is better*).

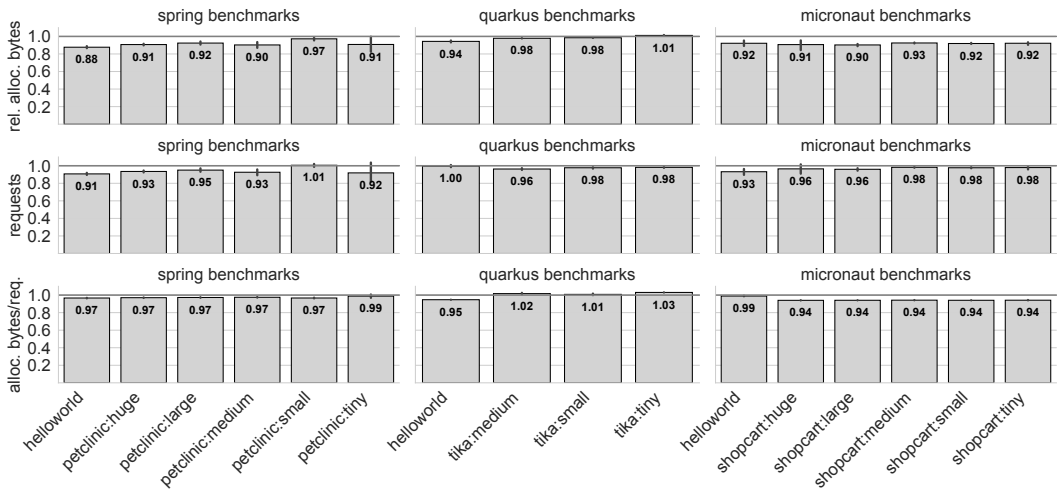

■ **Figure 7** Evaluation of our approach on microservice benchmarks, showing the ratio of the overall allocated bytes (*lower is better*), the number of performed requests (*higher is better*) and the overall allocated bytes per request (*lower is better*).

Due to the close connection between HashSet and HashMap (explained in Section 4), replacements of HashSet influence both the memory allocation counter for HashSet and for HashMap. For example, replacing a large HashSet with our custom memory-efficient implementation results in less memory allocated for HashMap, as the HashMap backing HashSet would not need to be allocated at all. Hence, we only list the memory allocation change for HashMap and HashSet together.

In *dacapo:pmd*, the number of bytes occupied by HashMaps, HashSets and their associated auxiliary objects is reduced by 77.95 %, for LinkedHashMaps by 0.87 %, and for ArrayLists by 14.18 %. The replacements of HashMaps and ArrayLists are particularly significant here, as these objects accounted for around 14.32 % and 29.48 % of all allocated bytes in the benchmark before the replacement. HashSets have no impact as they only account for 0.06 %. In total, our approach replaced 319 allocation sites in *dacapo:pmd*.

In *renaissance:scala-stm-bench7*, the number of bytes occupied by HashMaps, Hash-Sets, and associated auxiliary objects is reduced by 69.99 % In this case, almost all HashMap allocations came from HashSets. However, since we replace all HashSets, we automatically also prevent the HashMap allocations. This seems to be the main reason for the reduction, as HashMap allocations previously accounted for about 14.92 % of the total allocated memory. In this benchmark, we replaced 55 allocation sites.

In *renaissance:mnemonics* and *par-mnemonics*, our approach led to 52 and 53 replacements, respectively. Both benchmarks benefit from the HashSet replacements. In combination, HashSet and HashMap memory allocations were reduced by 26.65 % (mnemonics) and 26.07 % (par-mnemonics). While Section 7.1 suggests that *renaissance:fj-kmeans* would also be a potential target, analyzing its data structures shows why our approach has no real impact on the benchmark: ArrayLists are a major contributor (44.99 %) to the benchmark's allocated memory, but most are too large to be eligible for replacement and do not uniformly contain elements of primitive types.

The microservice benchmarks consist of two parts: The application under test, which in all cases is a webservice, and a benchmarking tool that generates a constant load in the form of HTTP requests. After warmup, the benchmark runs for a fixed amount of time during which the benchmarking tool performs as many requests as possible. Hence, the performance of the application is measured in successfully handled requests per second. This work focuses on the reduction in memory allocation achieved by our optimization. However, we need to consider the fact that our replacement data structures may be slower in certain cases. Thus, a direct comparison of the values for the total allocated bytes is not meaningful in a benchmark that does not have a fixed workload, as our replacement data structures might be the reason why fewer requests are handled and thus less work—that might cause additional object allocations—is performed. Therefore, we set the memory allocation in relation to the processed requests (*allocated bytes per request*). Figure 7 presents the results for the microservice benchmarks in terms of allocated bytes, requests, and allocated bytes per request with our replacements compared to the baseline.

With our replacements, we were able to reduce the allocated bytes per request by 2.94 % on average (geometric mean). Notably, we achieved memory allocation reductions in most microservice benchmarks. The *tika* microservice benchmarks are an outlier where the results indicate a regression. As the *rel. allocated bytes* metric shows, we can achieve a slight improvement in both *odt-medium* and *odt-small* workloads (the rather small impact is to be expected since data structures only account for about 1 % of the total allocated bytes). However, this improvement is offset by the lower number of requests that could be performed in the benchmark. For *odt-tiny*, in addition to the performance regression, we also see a 1.14 % increase in *rel. allocated memory*. This regression cannot be attributed to data structures, but rather to other objects that are created in the benchmark. This results in a regression of 3.02 % in terms of allocated bytes per request. Nevertheless, we were able to reduce the allocated bytes per request for HashMaps and HashSets by 36.01 %.

The *shopcart* benchmarks experience the biggest reduction in allocated bytes per request. In *shopcart:mixed-tiny*, for example, the number of bytes occupied by HashMaps, HashSets and their associated auxiliary objects is reduced by 73.45 %, for Linked-HashMaps by 11.01 %, and for ArrayLists by 29.10 % per request. In this benchmark, the replacements of HashMaps, LinkedHashMaps and ArrayLists are particularly important, as these types accounted for around 3.97 %, 3.67 %, and 5.26 % of all allocated bytes before the replacement.

The allocated memory in *quarkus:helloworld* is reduced by 5.27 %. Here, the allocated bytes for HashMaps/HashSets are reduced by 26.84 %, for LinkedHashMaps by 3.98 %, and for ArrayLists by 19.98 % per request. Particularly the HashMap replacements contribute to the reduction, as they originally accounted for around 18.62 % of the allocated memory.

**Impact of Partial Escape Analysis on Data Structure Replacement**   *Partial escape analysis* (PEA) [59] enables the compiler to prevent allocations of objects, including data structures. Therefore, the amount of data structures that can be replaced using our proposed optimization is reduced. However, as most compilers do not feature PEA, we decided to evaluate our approach with PEA disabled. The results show that we are able to reduce the allocated memory by 1.96 % for the standard benchmarks and by 7.04 % for the microservice benchmarks. The detailed results of applying our optimization without PEA can be found in Appendix F.

**Accuracy of Profiling and Evaluation of Fallbacks**   As some of our replacement data structures have a fallback mechanism that ensures that the implementation is semantically equivalent to the original data structure even if the profiling information is imprecise (e.g., if we insert an element into a SingletonHashMap), we also added an evaluation of the fallbacks that occurred in our benchmark set in Appendix D. There, we could see that aside from minor outliers in most benchmarks the fallback rate is as expected from our heuristics (below 5 %).

### 7.3  Impact on Performance and Binary Size

As our replacements are designed to minimize memory usage, they do have a performance impact on applications. However, standard benchmarks and microservice benchmarks only suffer from a minor performance penalty of 0.60 % and 4.19 %, respectively. Note that this slowdown is already considered in the results in Figure 7, as we normalized the allocated bytes by the number of performed requests.

As we introduce new types with new methods and therefore code that has to be included in the binary, we expect a minor increase in the size of the compiled binary executable. The evaluation indeed showed that our replacements have minimal impact on the size of the generated binary: Our approach increased the binary size by 0.36 % in standard benchmarks and by 0.66 % in microservice benchmarks. In Appendix E, we present charts that cover both performance and image size for each benchmark.

### 7.4  Discussion

The evaluation has shown that our approach is not generally beneficial for all applications: While we could achieve a reduction in allocated bytes by 1.63 % for standard benchmarks and by 2.94 % for microservice-based benchmarks, many benchmarks are largely unaffected as our target data structures do not contribute significantly to the total memory usage there. However, as shown in Section 7.2, benchmarks that do depend on data structures can experience improvements of up to 13.85 %. These results are mostly in line with our findings in Section 7.1. Furthermore, we could not identify major regressions in terms of memory usage. Since our approach (including all of the explained thresholds and heuristics) can be enabled and disabled via a simple command line switch we argue that it is a useful complement to Native Image and other similar compilers: Developers can use our optimizations if their applications largely depend on data structures and if memory optimization is a priority.

## 8   Related Work

There is an abundance of work dedicated to exploring usages of collections or containers as well as their improvement.

Most recently, Babynyuk [4] introduced data structure tuning via GraalVM Native Image. Their approach uses profiling information to optimize initialization parameters of data structures, such as HashMap's initial capacity. Their approach, however, is limited to swapping constructor arguments of HashMaps and they mention data structure replacement as a possible extension.

*Perflint* [31] instruments C/C++ standard library constructs to detect misuses of container components. It traces "state changes" of each data structure (e.g., a resize operation triggered after insertion) and subsequently uses a handcrafted cost model to predict inefficiencies in the code and report them to the developer. Jung et al. [26] go in a similar direction but use a machine learning model and hardware counters to get more accurate results and better predictions.

Xu et al. [71, 72] use bytecode instrumentation and static analysis to perform large-scale analyses for detecting *underutilized* or *overpopulated* Java Collections and making them visible to the developer. In contrast, our is work centered on automatically optimizing certain container allocation sites. While we base our optimizations on profiling, our approach is limited to information that we can detect at run time per allocation site with only limited contextual information. Their analysis most likely outperforms our heuristics as they also construct a call-graph for the detected container operations and therefore are able to infer a large number of optimizable containers. While they use mostly different benchmark sets compared to our approach, we see similar results on those that do overlap: Their work shows little to no findings for both *dacapo:xalan* and *dacapo:luindex* but could identify optimizable containers within *dacapo:pmd*. This also somewhat mirrors our evaluation results, as our approach showed little to no impact on the former two, while achieving significant improvements in *pmd*. Nevertheless, we think that extending our work with methods and metrics proposed by these authors can produce further interesting insights into data structures.

Costa et al. [12] provide a custom library that abstracts lists, sets, and maps that track properties of the corresponding collection per allocation site. During monitoring runs, the performance of the collection is tracked and implementations are swapped if a performance model suggests so. Additionally, they enable per-instance tracking and swapping, if an allocation site is revealed to create heterogeneously used collection instances. The abstract layer devised by Österlund and Löwe [44] works similarly— their data structures can swap their implementations internally—but they use a state machine that triggers transformation as soon as certain "biases" are exceeded. De Wael et al. [15] implemented *Just-in-Time Data Structures*, a library that enables run-time transformation of a data structure's internal representation. Users can add such data structures by defining *swap rules*.

In contrast to these works, our approach adapts the library classes themselves, without requiring an abstraction layer. Replacement is done during AOT compilation, hence neither data duplication nor run-time transformation is necessary.

De Sutter et al. [14] suggest using type constraints to determine valid replacements. They also use a points-to analysis and instrument collection classes for profiling. Their replacement process is based on using *sibling classes* and they propose a variety of optimizations, such as lazy initialization, removal or transformation of fields, or dedicated specializations for zero- or one-element collections (similar to our approach).

While the use of sibling types allows for greater customization of the replacement types, it is also potentially more error-prone, as all usages and accesses to the replaced collections have to be adapted as well. Similar to our approach, it could also lead to compliance issues when run-time reflection or type checks are used. Wang et al. [63] synthesize container replacements that use operations with lower time complexity than the existing ones. Therefore, they define method specifications for the Java Collections Framework that ensure preservation of the program behavior. Algorithmic complexity is not a major concern of our work as we currently focus on replacing data structures with "similar" implementations, but their insights could be helpful to extend our approach to even more data structures.

*Chameleon* by Shacham et al. [56] instruments the garbage collector to profile accurate metrics upon collection of a data structure. Each allocation of a collection is therefore replaced with a custom implementation, that wraps the underlying implementation. A rule engine then selects appropriate replacements based on the profiling information. This approach is interesting, as using wrappers *globally* could simplify the replacement process, as mainly the internal implementation would be swapped. There is certainly some overhead to this but—as they mention—maybe the virtual machine or compiler could mitigate that. Using Native Image's static analysis and closed-world assumption, this could be an interesting alternative for our approach instead of using subclasses.

## 9　Conclusion

In this paper, we presented an approach to extend GraalVM Native Image's PGO implementation by enabling instrumentation per allocation site, thereby tracking a multitude of metrics for data structures on allocation site granularity. We focused on profiling of (Linked)HashMaps, HashSets, and ArrayLists. We then performed an analysis of the size classes of the instances of these data structures and their usage in standard benchmarks and microservice benchmarks. In our analysis, we specifically highlight the prevalence of small or even empty data structures. Based on this analysis, we introduced both fixed-size and in general more memory-efficient implementations for each data structure. We propose to replace data structure allocations at allocation sites with replacement types based on the presented heuristics. Additionally, we presented a powerful approach to perform this replacement automatically. The approach is not limited to user code, but can also perform the replacement in third-party code. Our evaluation shows that memory allocation can be significantly reduced through our optimizations.

**Acknowledgements**　The authors Lukas Makor and Sebastian Kloibhofer contributed equally to the paper. This research project was partially funded by Oracle Labs. We thank all members of the Virtual Machine Research Group at Oracle Labs. We also thank all researchers at the Johannes Kepler University Linz's Institute for System Software for their support of and valuable feedback on our work.

**A**  **Replacement Analysis**

■ **Table 2** Number of replaced allocation sites per benchmark suite broken down by replacement type.

| Replacement Type | DA | SC | RE | SP | QU | MI |
|---|---|---|---|---|---|---|
| EmptyHashMap | 72 | 24 | 28 | 781 | 49 | 32 |
| SingletonHashMap | 25 | 1 | 18 | 213 | 28 | 33 |
| Size2HashMap | 14 | 3 | 6 | 91 | 31 | 22 |
| EconomicHashMap | 71 | 46 | 48 | 131 | 24 | 39 |
| EmptyLinkedHashMap | 4 | 0 | 0 | 246 | 4 | 60 |
| SingletonLinkedHashMap | 6 | 1 | 1 | 85 | 4 | 24 |
| Size2LinkedHashMap | 0 | 0 | 2 | 79 | 4 | 7 |
| EconomicLinkedHashMap | 0 | 0 | 0 | 42 | 0 | 27 |
| EmptyHashSet | 32 | 11 | 21 | 318 | 51 | 71 |
| SingletonHashSet | 17 | 2 | 4 | 226 | 10 | 35 |
| Size2HashSet | 7 | 0 | 0 | 120 | 24 | 26 |
| MemoryEfficientHashSet | 760 | 315 | 708 | 3436 | 1055 | 1263 |
| EmptyArrayList | 145 | 77 | 207 | 903 | 132 | 166 |
| SingletonArrayList | 96 | 28 | 99 | 700 | 97 | 170 |
| Size2ArrayList | 64 | 36 | 46 | 372 | 24 | 45 |
| IntArrayList | 3 | 0 | 1 | 5 | 0 | 12 |

DA ... dacapo, SC ... scala-dacapo, RE ... renaissance
SP ... spring, QU ... quarkus, MI ... micronaut

Table 2 lists the number of replaced allocation sites per benchmark suite, broken down by the corresponding replacement type. *Empty*, *Singleton*, and *Size2* refer to our fixed-size replacement types. *Economic(Linked)HashMap* refers to our memory-efficient (Linked)HashMap type and *MemoryEfficientHashSet* to our memory-efficient HashSet type.

All HashSets that are not replaced with specialized variants based on the profiling data, are replaced with our MemoryEfficientHashSet. Hence, the MemoryEfficientHashSet is the most frequently used replacement type across all benchmarks. As expected from our analysis, the size-0 replacement types are also frequently used. Interestingly, only our replacement type for primitive *int* ArrayLists (*IntArrayList*) is used (but still very rarely).

Figure 8 shows the allocation ratio of our replacement types relative to the original types (e.g., a value of 0.7 for ArrayList means that 70% of all allocated ArrayLists were actually of our replacement types). As we replace *all* HashSet allocations, we omitted these values from the figure.

The results for *renaissance:mnemonics* and *renaissance:par-mnemonics* are particularly interesting: As described before, these benchmark show high reductions, which must be largely due to HashSet, which were completely replaced, whereas (as we see in the chart) almost none of the other data structures were replaced. The reduction is largely due to *preventing* HashMap allocations via HashSet replacements.

The benchmark *renaissance:fj-kmeans* shows no replacements. While ArrayLists account for 44.99 % of all allocated bytes in this benchmark, most of them are too large to be considered for replacement.

Although *dacapo:pmd* is a benchmark that largely benefits from our replacements, we can still only replace 0.04 % of all HashMaps. This is due to the frequent use of `HashMap.entrySet` which prevents the replacement (see Section 3.1). We made similar observations in *renaissance:mnemonics*, *renaissance:par-mnemonics*, *renaissance:rx-scrabble*, and *renaissance:scrabble*.

In the microservice benchmarks, *micronaut:helloworld* stands out, as almost all of the targeted data structure allocations there were replaced. As expected, the different variants of the microservice benchmarks have similar allocations. Here too, we see that more than half of the data structure allocations could be replaced.

Figure 9 and Figure 10 show the distribution of our replacement data types. The chart depicts the relative percentage of all allocated replacement data structures that belong to the corresponding implementation. As an example, the HashMap chart (third row, first column) for *dacapo:pmd* shows that around 72.0 % of all allocations of replaced data structures belong to EmptyHashMaps, while 12.8 % are Singleton-HashMaps, 8.6 % Size2HashMaps, and 6.6 % EconomicHashMaps. In benchmarks without bars, e.g., LinkedHashMap replacements for *dacapo:avrora* no replaced data structures were allocated. Here, we see once again that among the fixed-size replacements our specialized implementations for empty data structures occur most often in most benchmarks. Outliers are LinkedHashMaps, where the standard benchmarks show hardly any replacements. In benchmarks that show replacements, mostly size-1 and size-2 replacements are used. The *quarkus* benchmarks show similar behavior— here the size-2 replacements dominate. The charts further show that the primitive specializations of ArrayLists see hardly any usage across all benchmarks.

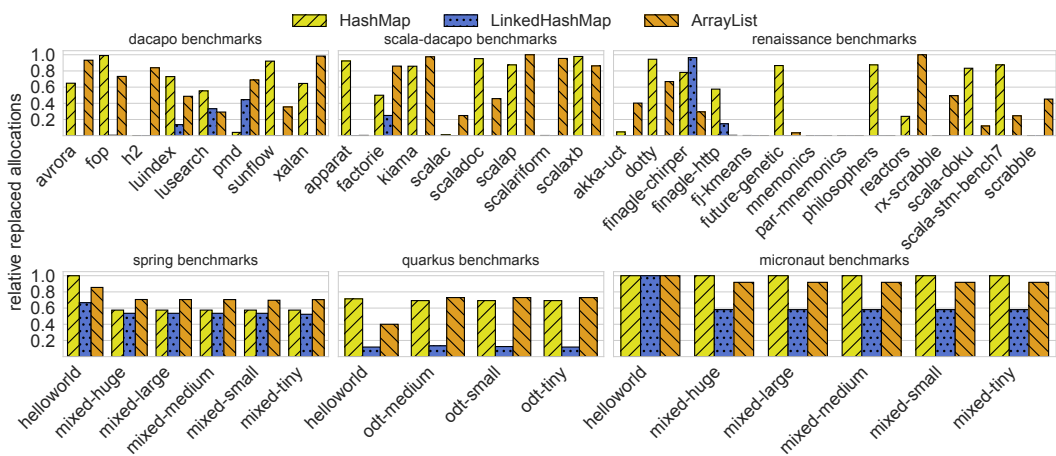

■ **Figure 8**  Ratio of data structures that were replaced in standard benchmarks and microservice benchmarks (*higher is better*). This evaluation excludes HashSets, as we replace all HashSet allocations.

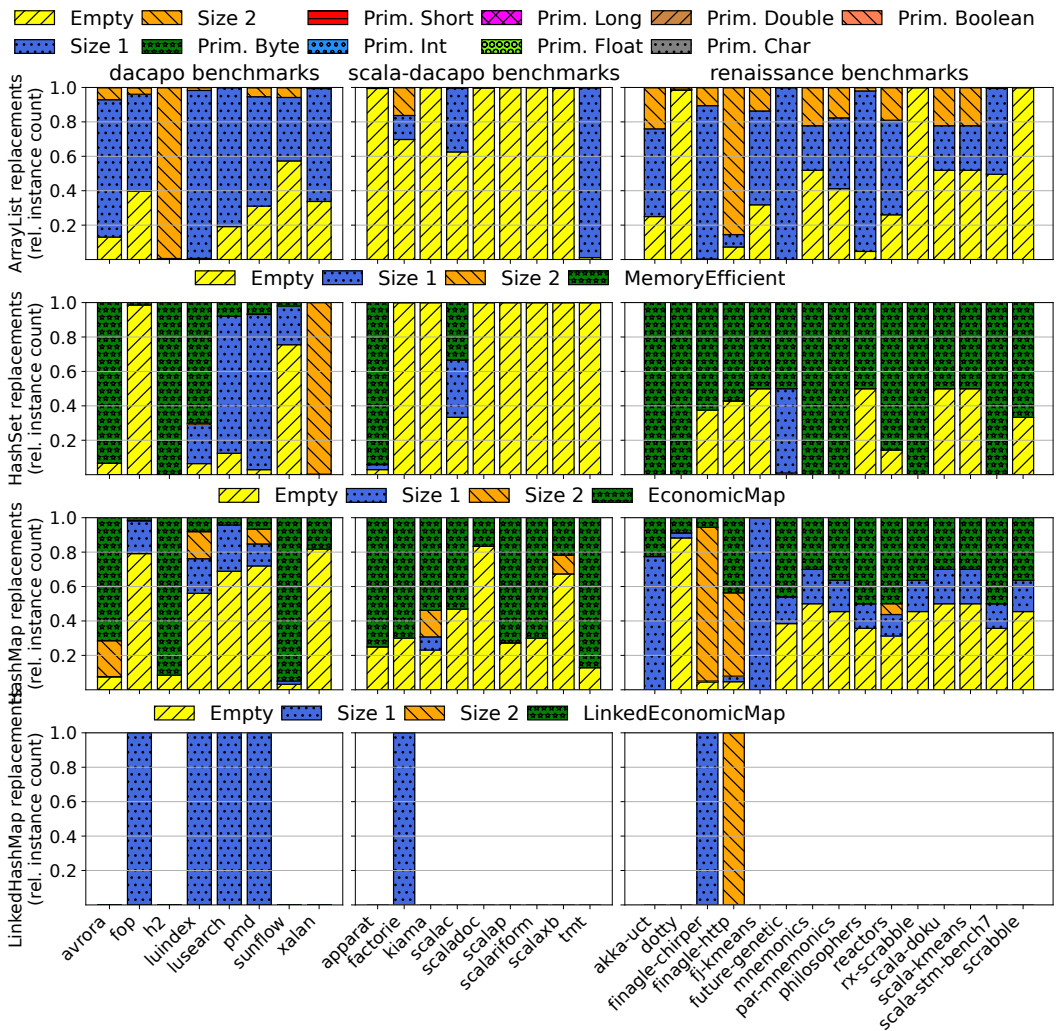

**■ Figure 9** Distribution of our specialized implementations (in terms of number of allocated instances) within all replacements in standard benchmarks.

**B** **Memory Allocation Evaluation Details**

Figure 11 shows the individual reductions in the overall allocated bytes per data structure (ArrayList, LinkedHashMap, and the combination of HashSet and HashMap as HashSet is backed by a HashMap in the original implementation). We achieve most reductions in the HashMaps and HashSets. Two factors come into play here: First, as shown in the introductory example, HashMaps cause bloat when filled with only few elements due to the relatively large minimal size and the backing Entry objects that store the actual data. Our fixed-size replacement as well as our EconomicMap variant cause some savings there. Second, HashSets are backed by HashMaps, thus the same weaknesses apply. In addition to fixed-size replacements, we also replace all remaining HashSet allocations with one that does not use a backing HashMap. This also reduces the memory usage per HashSet instance. On some occasions, the results suggests increases in the memory used by data structures, particularly LinkedHashMaps. These

**Automated Profile-Guided Replacement of Data Structures to Reduce Memory Allocation**

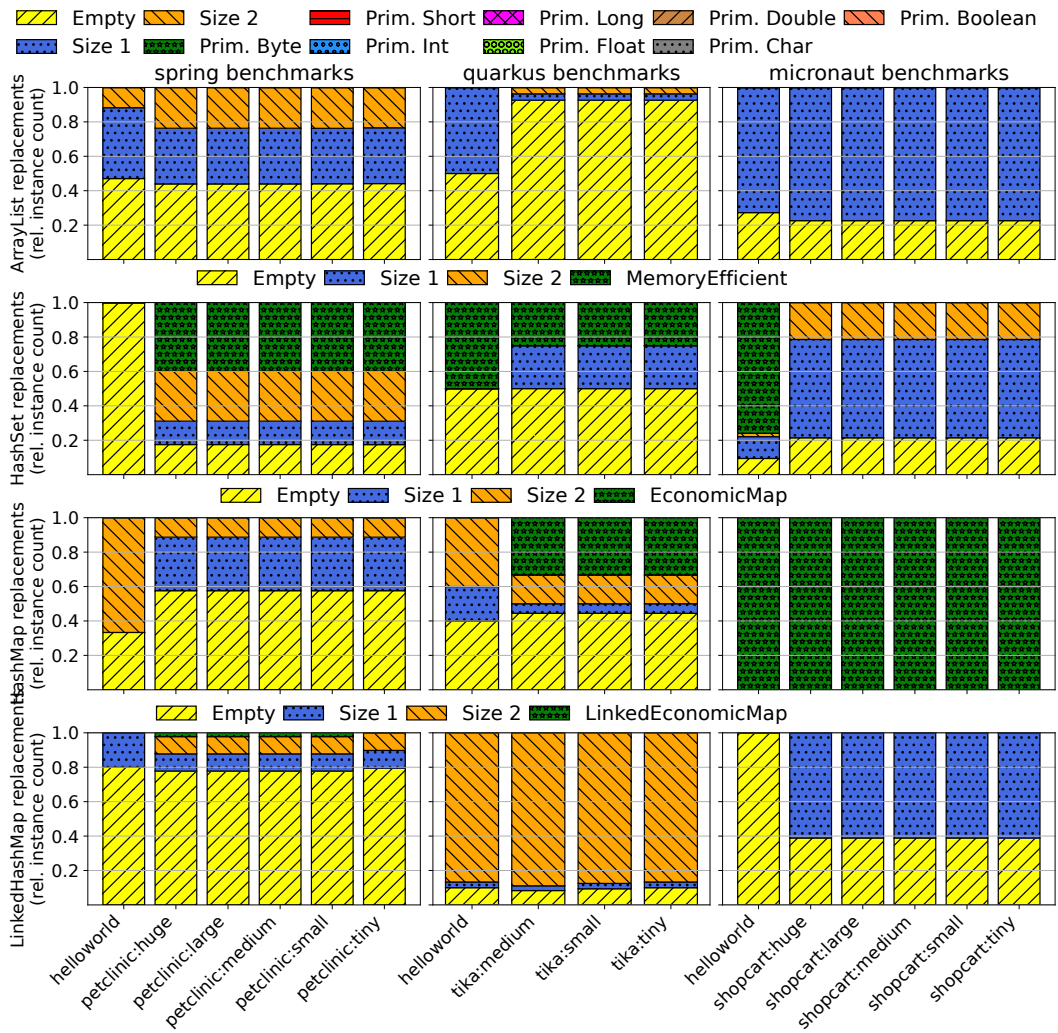

■ **Figure 10** Distribution of our specialized implementations (in terms of number of allocated instances) within all replacements in microservice-based benchmarks.

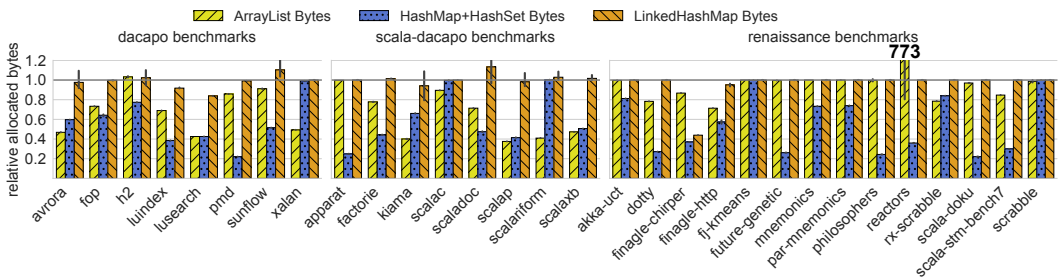

■ **Figure 11** Evaluation of our approach on standard benchmarks comparing the total allocated bytes per target data structure in the baseline (without replacement) with the allocated bytes after replacement (*lower is better*). This figure contains details corresponding to the evaluated allocated bytes from Figure 6 and the associated discussion. The outlier in *renaissance:reactors* is specifically marked as it exceeds the other bars significantly.

data points typically present themselves with large errors. We attribute this to the low number of replacements that occur in this category; in *scala-dacapo:scaladoc*, as an example, the number of allocated LinkedHashMaps vary between 3 and 4 (none of them replaced). We can see one interesting apparent regression in the allocated memory for ArrayLists in the *renaissance:reactors* benchmark. Here, the allocated bytes have increased significantly but the measurement error is also extremely high. In fact, in the baseline only around 700 ArrayLists are allocated. With our replacement, however, some benchmark executions suddenly show over 5 million ArrayList allocations. As the corresponding bar exceeds the chart, we specifically annotated the regression (773x). Investigation showed that in this non-deterministic case (hence the large error), our replacement negatively impacts the partial escape analysis optimization in the compiler. In the baseline, this optimization removes all ArrayList allocations at a certain allocation site, whereas this is prevented after our replacement.

While this is certainly a drawback of our approach, those ArrayLists are short-lived and typically collected during the first garbage collection; overall the benchmark still showed an improvement of 1.05 % in terms of allocated memory. However, the chart also shows that cases where our replacements lead to additional allocations occur only very rarely. Other than *reactors*, we could see no other benchmarks, where our impact on partial escape analysis caused a significant regression.

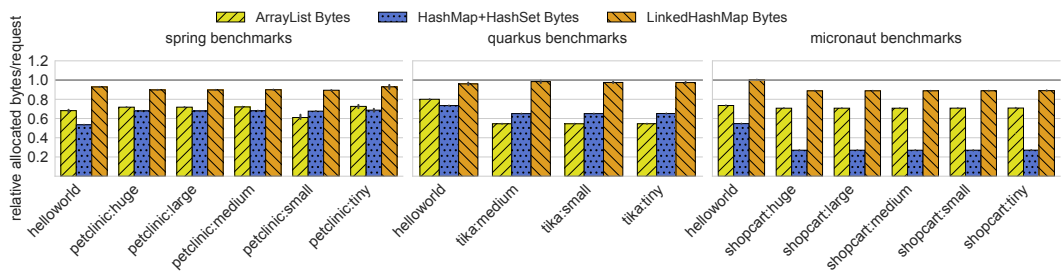

■ **Figure 12** Evaluation of our approach on microservice-based benchmarks, showing the ratio of the total allocated bytes per target data structure within each benchmark (*lower is better*). This figure contains details corresponding to the evaluated allocated bytes from Figure 7 and the associated discussion.

Figure 12 presents the same data for microservice-based benchmarks. While the results are similar across the different workload sizes (small, medium, etc.), we see the most reductions in the allocated bytes of ArrayLists and HashMaps/HashSets. Regarding the latter data structures, we contribute the reductions to the same observations as in the previous section about the results on standard benchmarks. Especially our replacements for empty ArrayLists and for those with 1 element are applied frequently in those benchmarks. While the numbers suggest potential for improvements in the *quarkus* benchmarks, the observed insignificance of data structures in those benchmarks as per Figure 5 explains the modest results in terms of overall allocated memory. The *quarkus:helloworld* benchmark is the only exception here, as it shows both reductions per data structure type and also promising results in the overall allocated bytes presented in Figure 7.

 **Lifespan Analysis of Replaced Objects**

To analyze how much of the long-lived objects are data structures, we instrumented the garbage collector to track objects that are promoted to the old generation. Objects that are stored in the old generation of the garbage collector typically survived multiple garbage collections and thus can be considered long-lived. Hence, our baseline value for this analysis is the number of bytes of all objects that were promoted to the old generation. Table 3 and Table 4 show the relative amount of memory in the old generation that is occupied by data structures that we target. Furthermore, we also list the ratio of data structures in the old generation that our approach replaced with a specialized implementation.

In the results for the standard benchmarks, we see on the one hand that there are quite many benchmarks were no significant amount of memory in the old generation is made up by data structures. However, there are also cases such as *renaissance:fj-kmeans*, *renaissance:pmd*, and *dacapo:avrora*, where data structures make up significant parts of the old generation memory. Unfortunately, as already discussed in Section 7.2 we cannot optimize most data structure allocations in *renaissance:fj-kmeans*, which also explains the low replacement rate of 0.2 %. For *renaissance:pmd* the high amount of data structures in the old generation is also in line with our investigation regarding the share of data structures of all allocated objects that we presented in Section 7.1. As *renaissance:pmd* is also the benchmark with the largest reduction in allocated bytes in our evaluation it is reasonable to expect that also in the old generation a high percentage of data structures can be replaced.

While the results of the old generation measurement for *dacapo:avrora* show that data structures make up a significant part (12.5 %) of the memory in the old generation and most of them (72.2 %) are replaced, we only see a minimal improvement in our evaluation in Section 7. The reason is that in *dacapo:avrora* data structures only make up a very small part of the total allocated objects.

The results for microservice-based benchmarks are less varied with most showing that at least some percentages of the memory in the old generation stem from data structures. In all benchmarks, we could replace a significant part of those long-lived data structures.

 **Fallback Evaluation**

As described in Section 4, we provide a number of replacement data structures that use a fallback mechanism to not violate the semantics of the original data structure implementation. One example thereof is an EmptyHashMap that is a specialized replacement for HashMaps that remain empty. If after replacement, however, the program behaves differently than in the programming run or if at a single allocation site a small portion of the allocated HashMaps do hold elements, the EmptyHashMap implementation still has to be able to also work with actual entries. In such cases, it therefore falls back to the original implementation. Naturally, this is a case that we want to avoid as much as possible as it eliminates the effective of the corresponding

■ **Table 3** Share of data structures in the garbage collector's old generation in terms of bytes (Old Gen DS Ratio) and ratio of data structure instances in the garbage collector's old generation that are replaced via our optimization (Old Gen Replaced DS Ratio) for standard benchmarks. The overall result is shown in the form of the arithmetic mean of the individual results.

| Suite | Benchmark | Old Gen DS Ratio | Old Gen Replaced DS Ratio |
|---|---|---|---|
| dacapo | avrora | 12.5 % | 72.2 % |
| | fop | 6.2 % | 51.5 % |
| | h2 | 4.4 % | 74.3 % |
| | luindex | 1.1 % | 75.2 % |
| | lusearch | 1.4 % | 55.4 % |
| | pmd | 15.4 % | 64.1 % |
| | sunflow | 0.0 % | 81.4 % |
| | xalan | 2.4 % | 67.5 % |
| scala-dacapo | apparat | 0.0 % | 52.6 % |
| | factorie | 0.0 % | 54.9 % |
| | kiama | 0.0 % | 83.8 % |
| | scalac | 0.0 % | 92.7 % |
| | scaladoc | 0.0 % | 81.6 % |
| | scalap | 2.5 % | 79.6 % |
| | scalariform | 4.9 % | 11.5 % |
| | scalaxb | 0.0 % | 78.0 % |
| | tmt | 0.0 % | 81.4 % |
| renaissance | akka-uct | 0.0 % | 1.5 % |
| | dotty | 0.1 % | 49.5 % |
| | finagle-chirper | 0.5 % | 12.7 % |
| | finagle-http | 0.1 % | 29.6 % |
| | fj-kmeans | 51.6 % | 0.2 % |
| | future-genetic | 0.1 % | 0.5 % |
| | mnemonics | 1.6 % | 6.7 % |
| | par-mnemonics | 2.7 % | 6.7 % |
| | philosophers | 0.0 % | 26.3 % |
| | reactors | 0.0 % | 20.0 % |
| | rx-scrabble | 6.4 % | 87.0 % |
| | scala-doku | 0.0 % | 62.5 % |
| | scala-kmeans | 0.0 % | 62.5 % |
| | scala-stm-bench7 | 0.0 % | 99.7 % |
| | scrabble | 1.1 % | 0.0 % |
| | **Overall** | **3.6 %** | **50.7 %** |

DS ... data structures

■ **Table 4** Share of data structures in the garbage collector's old generation in terms of bytes (Old Gen DS Ratio) and ratio of data structure instances in the garbage collector's old generation that are replaced via our optimization (Old Gen Replaced DS Ratio) for microservice benchmarks. The overall result is shown in the form of the arithmetic mean of the individual results.

| Suite | Benchmark | Old Gen DS Ratio | Old Gen Replaced DS Ratio |
|---|---|---:|---:|
| spring | helloworld | 0.4 % | 96.4 % |
| | petclinic:huge | 3.1 % | 86.6 % |
| | petclinic:large | 2.7 % | 89.5 % |
| | petclinic:medium | 2.5 % | 90.0 % |
| | petclinic:small | 2.4 % | 90.1 % |
| | petclinic:tiny | 0.7 % | 98.3 % |
| quarkus | helloworld | 2.7 % | 75.1 % |
| | tika:medium | 2.2 % | 54.9 % |
| | tika:small | 4.3 % | 51.3 % |
| | tika:tiny | 7.3 % | 45.8 % |
| micronaut | helloworld | 3.6 % | 95.7 % |
| | shopcart:huge | 6.8 % | 71.5 % |
| | shopcart:large | 6.2 % | 72.0 % |
| | shopcart:medium | 6.6 % | 71.9 % |
| | shopcart:small | 6.5 % | 71.9 % |
| | shopcart:tiny | 5.9 % | 70.4 % |
| | **Overall** | **4.7 %** | **59.4 %** |

DS ... data structures

replacement data structure and produces overhead in terms of allocated memory. While our replacement heuristics are designed to minimize those cases, we conducted an evaluation that shows how often fallbacks still occur in each benchmark. Table 5 and Table 6 contain the results of this evaluation for standard and microservice-based benchmarks, respectively. The allocations correspond to the number of instances of replacement data types. Hence, it does not include allocations of the original type (e.g., it counts EmptyHashMap but not HashMap allocations). We can see that the results are varying: While most benchmarks (particularly the *spring*, and *micronaut* benchmarks) show hardly any fallbacks compared to the overall allocations, there are some outliers. *scala-dacapo:factorie* has a fallback rate of around 8 % as in each run of the benchmark only around 62 instances of replaced data structures were allocated, 5 of which typically used the fallback. *dacapo:h2* shows an even higher fallback rate of nearly 18 % of the replaced data structure instances. In this case, the most fallbacks stem from the Size2ArrayList implementation, where we see around 1.5 million fallbacks against around 9 million allocations. However, in the majority of cases, the fallbacks are below 1 % of the allocations.

■ **Table 5** Evaluation of the fallbacks that occurred in standard benchmarks compared to the total allocations of replaced data structures. As an example, 47395 allocations and 21 fallbacks for *dacapo:luindex* mean that out of 47395 allocations of replaced data structures that have a fallback mechanism (SingletonHashMap, IntArrayList, EconomicMap, etc.) 21 had to resort to the fallback as the profiling information was incorrect.

| Suite | Benchmark | Allocations | Fallbacks (%) |
|---|---|---|---|
| dacapo | avrora | 1,075 | 0 (0.00 %) |
| | fop | 5,068,944 | 74,218 (1.46 %) |
| | h2 | 8,933,134 | 1,578,389 (17.67 %) |
| | luindex | 47,395 | 21 (0.04 %) |
| | lusearch | 18,236,094 | 0 (0.00 %) |
| | pmd | 266,855,089 | 4,097,859 (1.54 %) |
| | sunflow | 2,573 | 0 (0.00 %) |
| | xalan | 610,356 | 116 (0.02 %) |
| scala-dacapo | apparat | 5,661 | 0 (0.00 %) |
| | factorie | 62 | 5 (8.06 %) |
| | kiama | 6,576,480 | 170,241 (2.59 %) |
| | scalac | 180,364 | 1,974 (1.09 %) |
| | scaladoc | 363,812 | 42 (0.01 %) |
| | scalap | 303,707 | 0 (0.00 %) |
| | scalariform | 180,228 | 1,932 (1.07 %) |
| | scalaxb | 1,400,688 | 0 (0.00 %) |
| | tmt | 29,935 | 0 (0.00 %) |
| renaissance | akka-uct | 85,563 | 491 (0.57 %) |
| | dotty | 752,634 | 11,600 (1.54 %) |
| | finagle-chirper | 10,818,215 | 54,218 (0.50 %) |
| | finagle-http | 11,690 | 653 (5.59 %) |
| | fj-kmeans | 30,057 | 0 (0.00 %) |
| | future-genetic | 500,257 | 50 (0.01 %) |
| | mnemonics | 16,981,703 | 0 (0.00 %) |
| | par-mnemonics | 16,981,951 | 0 (0.00 %) |
| | philosophers | 313 | 0 (0.00 %) |
| | reactors | 296 | 3 (1.01 %) |
| | rx-scrabble | 8,313,160 | 0 (0.00 %) |
| | scala-doku | 39 | 0 (0.00 %) |
| | scala-kmeans | 39 | 0 (0.00 %) |
| | scala-stm-bench7 | 1,053,113 | 0 (0.00 %) |
| | scrabble | 401,142 | 0 (0.00 %) |

■ **Table 6** Evaluation of the fallbacks that occurred in microservice-based benchmarks compared to the total allocations of replaced data structures.

| Suite | Benchmark | Allocations | Fallbacks (%) |
|---|---|---|---|
| spring | helloworld | 40,042,641 | 171 (0.00 %) |
| | petclinic:huge | 179,811,902 | 269,880 (0.15 %) |
| | petclinic:large | 180,675,243 | 270,772 (0.15 %) |
| | petclinic:medium | 139,226,377 | 208,837 (0.15 %) |
| | petclinic:small | 81,391,457 | 123,503 (0.15 %) |
| | petclinic:tiny | 46,552,227 | 73,289 (0.16 %) |
| quarkus | helloworld | 15,133,461 | 6 (0.00 %) |
| | tika:medium | 21,769,018 | 863,190 (3.97 %) |
| | tika:small | 14,798,109 | 601,849 (4.07 %) |
| | tika:tiny | 8,891,480 | 362,912 (4.08 %) |
| micronaut | helloworld | 23,961,085 | 23 (0.00 %) |
| | shopcart:huge | 114,206,335 | 70 (0.00 %) |
| | shopcart:large | 120,695,207 | 74 (0.00 %) |
| | shopcart:medium | 62,607,351 | 71 (0.00 %) |
| | shopcart:small | 32,749,425 | 71 (0.00 %) |
| | shopcart:tiny | 14,671,426 | 67 (0.00 %) |

## E    Performance and Binary Size Evaluation

We also evaluated the impact of our approach in terms of run-time performance and binary size. Figure 13 shows these results for standard benchmarks, where we see that most benchmarks exhibit minor slowdowns (0.60 % on average). One outlier is *scala-dacapo:kiama*, where our replacements result in an improvement of 20.86 %. Similarly, *renaissance:rx-scrabble* and *renaissance:fj-kmeans* are positively impacted. The renaissance benchmarks *mnemonics* and *par-mnemonics* show the most significant regression with slowdowns of 22.98 % and 18.74 %, respectively. Both benchmarks make heavy use of Java Streams and recursion. Unfortunately, our replacement data structures (particularly full HashMap and HashSet replacements) are not optimized for those use cases: For HashMap, streams typically require materialization of entry objects (see Section 3.1) and for others, native iterators are typically highly optimized for efficient and fast iteration. In contrast, our replacement data structures are specifically optimized for memory efficiency. We therefore deliberately omitted some features of the original data structures such as the caching of created *view objects* (e.g., the set of the keys for a HashMap). HashMap also has a feature that avoids long collision chains and the accompanying performance impact by restructuring the nodes internally as a tree [7]. Our replacement data structures lack such a feature. There is certainly further potential for optimizing the run time of our replacement data structures.

Figure 14 shows the impact of our approach on the binary size for microservice-based benchmarks. Most benchmarks are unaffected, only the shopcart variants suffer from a minor increase in the binary size. As we introduce new types into the compilation, a slight increase in the binary size is to be expected, but as the charts show, these effects are mostly insignificant.

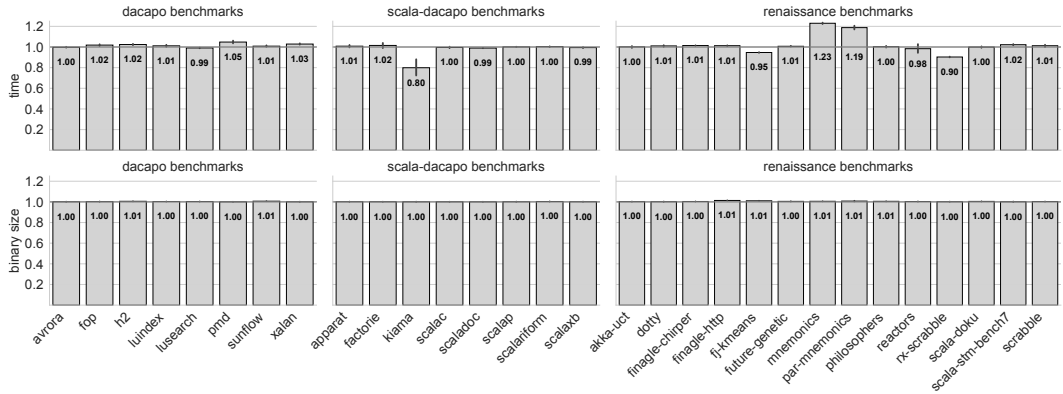

■ **Figure 13** Impact of our approach on the run time and binary size of the standard benchmarks (*lower is better*).

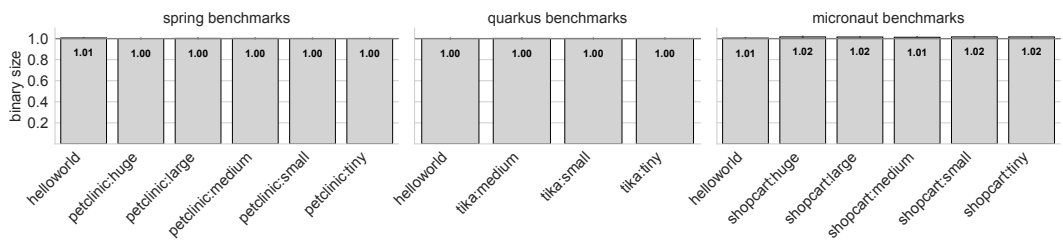

■ **Figure 14** Impact of our approach on the binary size of the microservice benchmarks (*lower is better*).

## F  Evaluation without Partial Escape Analysis

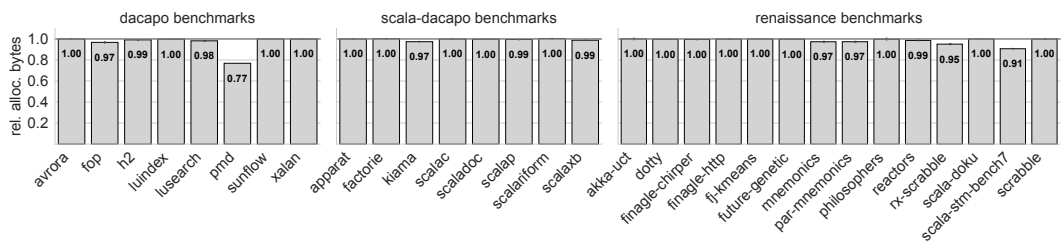

■ **Figure 15** Impact of our approach on standard benchmarks with PEA disabled. The chart shows the ratio of the total allocated bytes relative to the baseline (*lower is better*).

*PEA* and *scalar replacement* [59] are key optimizations in the GraalVM compiler. During compilation, they traverse the Graal IR graph and remove allocations of objects

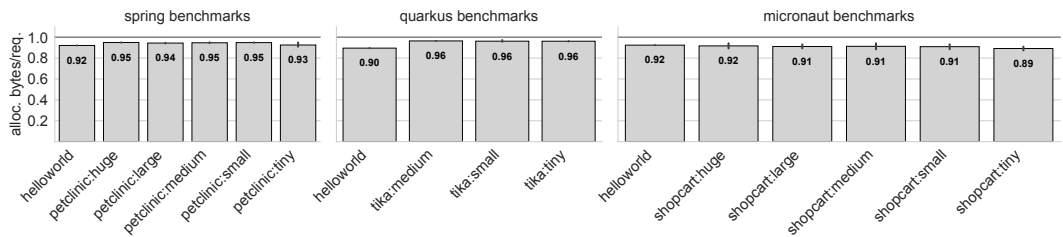

**■ Figure 16** Impact of our approach on microservice benchmarks with PEA disabled. The chart shows the ratio of the total allocated bytes (per request) relative to the baseline (*lower is better*).

**■ Table 7** Left: number of benchmarks that led to more replacements with PEA versus without PEA. Right: number of benchmarks that led to a higher ratio of replacements with PEA versus without PEA.

| Data Structure | #Benchmarks with Higher Replacement Count | | #Benchmarks with Higher Replacement Ratio | |
|---|---|---|---|---|
| | w/ PEA | w/o PEA | w/ PEA | w/o PEA |
| HashMap | 2 | 11 | 3 | 11 |
| LinkedHashMap | 0 | 27 | 0 | 27 |
| HashSet | 0 | 3 | 0 | 0 |
| ArrayList | 0 | 22 | 11 | 9 |

that do not "escape" the current context—i.e., objects that are not returned from the current method, are not stored in static fields or fields of other "escaping" objects. Particularly *partial* escape analysis (in contrast to regular escape analysis) enables such optimizations even over branching code by delaying and moving allocations, thus greatly extending their applicability [59]. Currently, however, PEA is not a very widely used optimization; even the HotSpot C2 compiler lacks this particular enhancement [22]. Since we argue that our approach is applicable to different compilers that support PGO, we also evaluated it under the assumption that PEA is not available.

In this evaluation, we saw an average reduction of the allocated bytes by 1.96 % on standard benchmarks and an average reduction of the allocated bytes, normalized by the number of requests, by 7.04 % on microservice benchmarks. Particularly the latter results significantly outperform the ones with PEA enabled (a reduction of 2.94 %).

Table 7 lists the number of benchmarks where the absolute number and the percentage of replaced instances are higher without PEA. We consider results that differ by less than 1 % as equal (hence they are not included in the table), as such small differences are within the margin of error of the measurements. Note that HashSet replacement percentages are expected to be equal since we always replace all allocated HashSets. From these results, we conclude that more data structures can be replaced without PEA than with PEA (both in terms of absolute and relative numbers). This makes our data structure replacement even more attractive for compilers that do not feature PEA.

## About the authors

**Lukas Makor** is a PhD student at the Johannes Kepler University in Linz, Austria. Contact him at lukas.makor@jku.at.
https://orcid.org/0000-0003-4683-9824

**Sebastian Kloibhofer** is a PhD student at the Johannes Kepler University in Linz, Austria. Contact him at sebastian.kloibhofer@jku.at.
https://orcid.org/0000-0001-5630-2372

**Peter Hofer** is a researcher at Oracle Labs. Contact him at peter.hofer@oracle.com
https://orcid.org/0009-0005-4725-1514

**David Leopoldseder** is a researcher at Oracle Labs. Contact him at david.leopoldseder@oracle.com
https://orcid.org/0000-0003-4683-9824

**Hanspeter Mössenböck** is a professor at the Johannes Kepler University in Linz, Austria. Contact him at hanspeter.moessenboeck@jku.at
https://orcid.org/0000-0001-7706-7308